\documentclass[12pt]{iopart}

\usepackage{array}
\usepackage{graphicx}
\usepackage{epstopdf}
\usepackage{iopams}  
\usepackage[latin1]{inputenc}
\usepackage[T1]{fontenc}
\usepackage[english]{babel}
\usepackage{lmodern}
\usepackage{siunitx}
\usepackage[parfill]{parskip}
\usepackage{multirow}

\begin{document}

\title{Optical dipole micro-trap for atoms based on crossed planar photonic waveguides}

\author{Yuri B. Ovchinnikov, Folly Eli Ayi-Yovo, and Alessio Spampinato}

\address{National Physical Laboratory, Hampton Road, Teddington TW11 0LW, UK}
\ead{yuri.ovchinnikov@npl.co.uk}
\vspace{10pt}
\begin{indented}
\item[]30 September 2023
\end{indented}

\begin{abstract}
Optical dipole micro-traps for atoms based on constructive superposition of two-colour evanescent light waves, formed by corresponding optical modes of two crossed suspended photonic rib waveguides, are modelled. The main parameters of the traps for rubidium atoms, such as potential depth, tunnelling rates of atoms from the trap and coherence time of the trapped atoms are estimated.   

\end{abstract}

\section{Introduction}

Optical dipole traps for ultra-cold atoms \cite{1} are widely used in modern experimental atomic physics. Most such traps are formed by single or several focused laser beams, which are propagating in a vacuum. These traps are relatively simple and can be considered conservative traps on time scales of transition of the trapped atoms to higher energy levels with subsequent spontaneous emission of photons. This makes it possible to use such traps for coherent manipulations with atoms, including all-optical production of atomic Bose-Einstein condensates (BEC) \cite{2}. 

The first proposal of a surface optical dipole trap for atoms \cite{3} formed by two-colour evanescent light waves of different penetration depths was based on total internal reflections of two focused laser beams from a surface of a glass prism. The main advantage of the approach is that the angles of reflection of the laser beams can be very close to the critical angles of the total internal reflection, which provide a relatively large penetration depth of the corresponding evanescent light waves into a vacuum. It makes it possible to move the minimum of the surface dipole trap far enough from the prism surface to reduce the influence of Casimir-Polder attraction of atoms to the dielectric surface of the prism. An experimental demonstration of such a trap for caesium atoms \cite{4}, where an evaporative cooling of atoms in the trap to a phase-space density of 0.1 was demonstrated. The same principle can be used for setting up two-dimensional surface optical lattices \cite{5}, which has not been demonstrated experimentally to date. It looks also quite logical to use optical waveguides, which support the simultaneous propagation of two optical modes of different colours, to trap or guide atoms in the corresponding two-colour evanescent light waves generated near the surfaces of such waveguides. So far such trapping of atoms was realised only for optical nanofibres \cite{6}. 
There were also several proposals of subwavelength lattices for atoms based on a combination of optical dipole forces and short-range interaction forces between atoms and dielectric or metal nanostructures \cite{7,8,9}, although it is not clear how to download atoms in such lattices.

On the other hand, such an evanescent wave trapping (EWT) of atoms next to planar optical waveguides was never demonstrated despite multiple proposals \cite{10,11,12,13,14}. There are two main reasons why the EWT did not work for planar waveguides. First, all waveguides used a guiding dielectric layer on top of a dielectric substrate of a smaller refractive index. In that case, the maximum propagation angle of the optical mode of the guiding layer is limited by the angle of total internal reflection at the interface between two dielectric media, which is smaller than the angle of total internal reflection in a dielectric waveguide surrounded by vacuum only, as takes place in optical nanofibres. Therefore, the penetration depths in a vacuum of the evanescent light waves of propagating modes of planar optical waveguides on dielectric substrates are always essentially smaller compared to optical nanofibres. This leads to the necessity of using very large laser powers in the corresponding optical modes of such planar waveguides to overcome the attractive Casimir forces near the surface of the waveguide. Second, in a planar configuration of a waveguide, the intensities of evanescent light waves are decreasing at the side edges of the waveguide, where the attractive Casimir force becomes dominant. This makes the lateral depth of the corresponding EW optical dipole potential much smaller than its depth in the direction, which is orthogonal to the surface of the waveguide \cite{15}. We will call that effect an asymmetry of the EWT potential for planar waveguides.

Recently we have shown that the problem of small penetration depth of the evanescent light waves in planar waveguides can be resolved with suspended optical waveguides \cite{15}. The asymmetry of the EWT potential in that case still exists, which is compromising its total depth. Therefore, the corresponding EWTs should be operated with colder atoms.
Such a suspended membrane rib (or ridge) waveguide based on aluminium oxide, which was intentionally designed for atomic trapping applications, was demonstrated and tested \cite{16}. It is remarkable, that this suspended waveguide has rather low losses in visible and NIR optical spectrum and can withstand powers of the guiding light up to 30 mW, which is more than enough for corresponding optical trapping and guiding of atoms. Another advantage of this approach is that it provides very good optical access to the waveguide due to the large total area of the membrane, which is supposed to provide convenient loading of the waveguide with ultra-cold atoms from any external atomic trap, like magneto-optical traps (MOT) or optical dipole traps.

In this paper we consider two-colour evanescent light wave micro-traps formed by two crossed suspended SiO$_2$ rib waveguides. 
The paper is organised as follows. In the second section calculations of optical fields in crossed suspended optical rib waveguides are presented. The third section presents the modelling of the trapping potentials of the micro-traps based on the crossed photonic waveguides and the main parameters of these traps are estimated. Finally, in the conclusion section, we summarise the results of our investigations and provide an outlook on possible applications of such micro-traps.

\section{Modelling of optical fields of a trap based on crossed suspended optical rib waveguides}

\subsection{Design and principle of the trap}

The idea of a crossed optical dipole trap (ODT), which is usually formed by two crossed Gaussian laser beams, the frequencies of which are detuned to the red side of the main atomic transition, is rather old \cite{17}. The main advantage of the crossed ODT over an ODT based on a single focused laser beam \cite{18} is it provides more even trapping strength in all directions, while its total volume can be larger. That is why the crossed ODT are widely used for an all-optical production of atomic Bose-Einstein condensates \cite{2,19}, where good compression of atoms in ODTs and their large volumes are important.

\begin{figure}[h]
	%\begin{verbatim}
	\centering
	\includegraphics[scale=0.5]{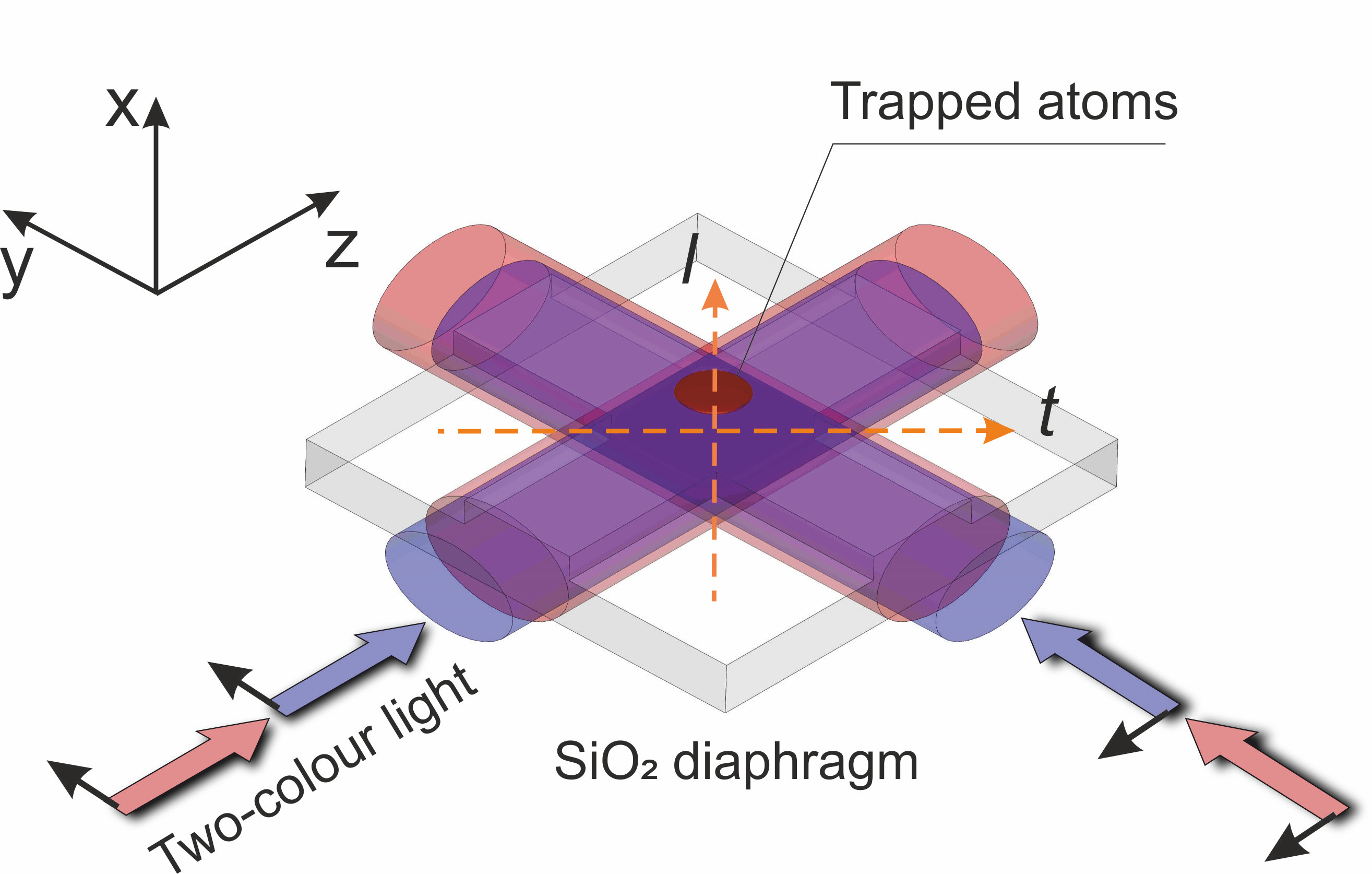}
	\label{fig1}
	\caption{Schematics of a two-colour evanescent optical dipole trap for atoms based on two crossed suspended optical rib waveguides.}
	%\end{verbatim}
\end{figure}

That principle of formation of tight optical dipole traps by crossed laser beams can be naturally extended to evanescent wave traps based on crossed planar waveguides. Figure 1 shows two suspended optical rib waveguides, which are crossed at the right angles. It is supposed that the waveguides are identical and each of them supports single-mode propagation of two different colour modes, which are necessary for guiding atoms in optical dipole potentials formed by corresponding evanescent light fields \cite{15}. Added to each other potentials of the two crossed atomo-photonic waveguides \cite{16} form an optical dipole trap in the cross region of the waveguides.
There is one essential difference between the crossed waveguide evanescent wave optical dipole traps (CWEWT) and the crossed ODTs based on Gaussian laser beams crossed in free space. Optical modes of CWEWTs are expected to be perturbed at the cross region due to different characters of guided light propagation in that region.
 
\subsection{Numerical simulation of evanescent light fields of the trap}

We consider two suspended optical rib waveguides, same as above, which are crossed to each other at the right angle, at it is shown in Figure 1. The basic configuration of suspended optical rib waveguides consists of a SiO$_2$ membrane (slab) with thickness $h=300$ nm and a rib of height $h_{rib}=15$ nm and width $w_{rib}=2$ $\mu$m \cite{12}. Our first configuration includes $\lambda_{b1}=720$ nm and $\lambda_{r1}=850$ nm modes, which propagate along the two suspended optical rib waveguides in single-mode regimes. The modes are far-frequency-detuned to the blue and red sides of the main transition ($\lambda_0=780$ nm) of the rubidium atom correspondingly. Initially, a propagation of one optical mode along just one of the two waveguides, which is oriented along $y$-axis, was simulated. In all of these numerical simulations, we used Lumerical FDTD software.

 \begin{figure}[h] 
 	\centering
 	\includegraphics[scale=0.6]{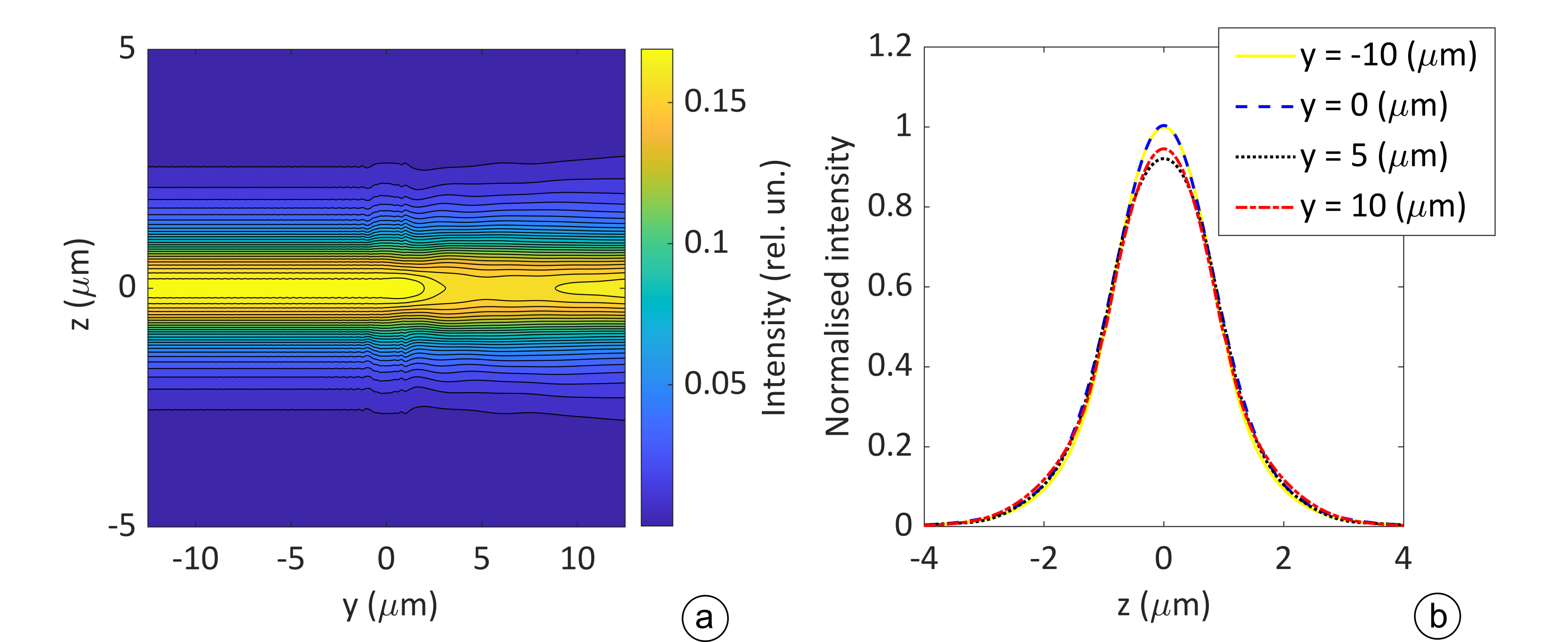}
 	\caption{a) Intensity distribution of the $\lambda_{r1}=850$ nm TE00 mode of the suspended optical rib waveguide with $w_{rib}=2$ $\mu$m width of the rib, which is crossing another identical waveguide at right angle at $y$=0; b) Cross sections of the intensity distribution along $z$-axis at $y =-10, 0, 5, 10$ $\mu$m.}
 	\label{fig2}

 \end{figure}

The cross region between the two waveguides can be considered as a gap between the two parts of the waveguide, where the propagating light is not confined in the transverse direction (along the $z$-axis). It is expected that the optical mode of that rib waveguide will experience a 2D diffraction at the cross region. That diffraction is quite similar to the emission of light from a single-mode optical fibre in a free space, which can be well approximated by the propagation of a Gaussian laser beam, the waist of which is located at the facet of a waveguide or an optical fibre. Therefore, the 2D-divergence of the propagating mode in the cross region can be characterised by a Rayleigh length, which is equal to $z_R^m=\pi w_m^2 n_m^{eff}/\lambda_m$, where $w_m$  is the lateral radius of propagating mode in the optical rib waveguide, $n_m^{eff}$ is effective refractive index \cite{20} for the mode propagating in the cross region and $\lambda_m$ is a wavelength of the mode's light in vacuum. For example, at $\lambda_{b1}=720$ nm, $w_{b1}=1.646$ $\mu$m and $n_{b1}^{eff}=1.386$, the corresponding Rayleigh length is equal to $z_R^{b1}=16.2$ $\mu$m, which is essentially larger than the size of cross region $w_{rib}=2$ $\mu$m. For the red frequency detuned mode $\lambda_{r1}=850$ nm, the Rayleigh length is $z_R^{r1}=20.5$ $\mu$m, taking into account $w_{r1}=2.03$ $\mu$m and $n_{r1}^{eff}=1.347$. Therefore, the degrees of diffraction for the two modes at the cross region are comparable to each other. An additional expected effect is a partial reflection of the mode from the interface between the waveguide and the cross region, which are characterised by slightly different effective refractive indexes. 

Figure 2 shows the simulated spatial intensity distribution of 850 nm light at the surface of the waveguide. The initial light distribution corresponds to the TE00 mode of the rib waveguide, the polarization of which lies in the $y0z$-plane of the membrane (Figure 1). For the 720 nm TE00 mode, the corresponding intensity distribution is very similar. It shows that the propagating optical modes are slightly perturbed at the cross region, from $y=-1$ $\mu$m to $y=1$ $\mu$m. One can see, that even after the cross region the light field of the mode is slightly expanded in the transverse direction, which is accompanied by a slight decrease in the optical field intensity at the centre of the waveguide.
Although, the intensity at the centre of the mode is getting restored at $y\gtrsim 10$ $\mu$m. Magnitudes of the 850 nm mode perturbations in its centre can be also estimated from Figure 5, which shows intensity distributions of two such crossed and non-interfering modes.
 
 \begin{figure}[h]
 	\centering
 	\includegraphics[scale=0.6]{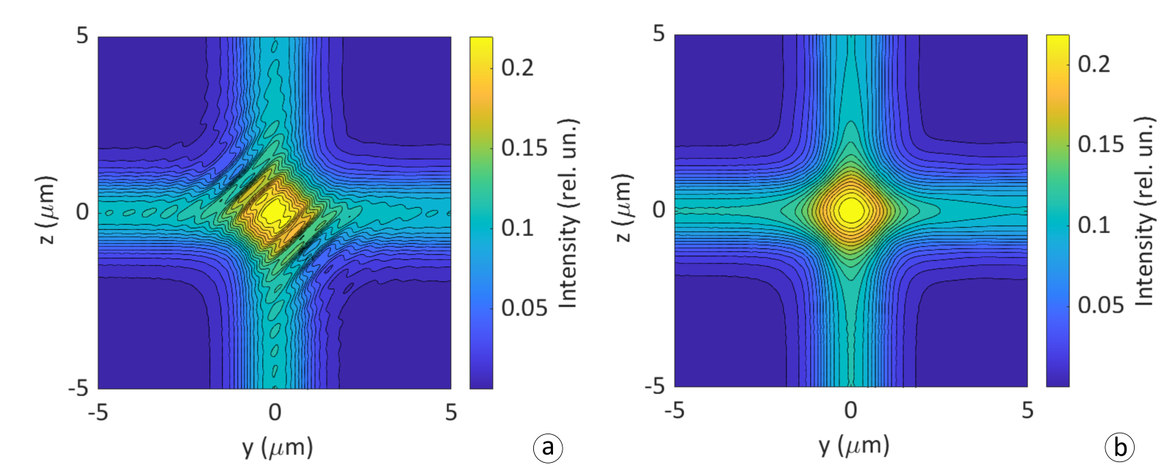}
 	\label{fig3}
 	\caption{a) Intensity distribution in the two crossed waveguides for 720 nm light modes in presence of interference between the two modes in the cross region; b) Intensity distribution in absence of interference between the two 720 nm modes.}
 \end{figure}
 
 Two optical modes of the same phase and frequency, which simultaneously propagate along the two crossed waveguides, experience partial interference in the cross region. A corresponding intensity distribution of light at the surface of crossed waveguides for $\lambda_{b1}=720$ nm modes is shown in Figure 3a. The period of observed interference fringes along the $t$-diagonal (see Figure 1) with period $d_{b1}=371$ nm in the cross region (Figure 3a, Figure 4) agrees with the effective index of refraction $n_{b1}^{eff}=1.386$ for this mode in the cross region. Note, that the interference takes place despite the initial polarisations of the two crossed modes being orthogonal to each other. The observed interference fringes can be explained by the diffraction divergence of the orthogonal modes in the cross region. This explains also why the relative amplitude of the interference fringes is close to zero at the centre ($t$=0) and is rising to $15\%$ at the wings of distribution, where polarisations of the diverged optical fields are not exactly orthogonal to each other. As an additional test of the fact that these intensity fringes are related to interference between the two modes, relative phase fluctuations of the two modes were introduced. Indeed the time-averaged phase fluctuations lead to the washing out of the interference fringes. Such a periodic intensity modulation of the evanescent light waves leads to the fragmentation of the optical dipole trap in the cross region into an array of smaller traps, which can be considered miniature optical lattices. However, it is not exactly what we want to achieve.   
 
\begin{figure}[h]
	\centering
	\includegraphics[scale=0.5]{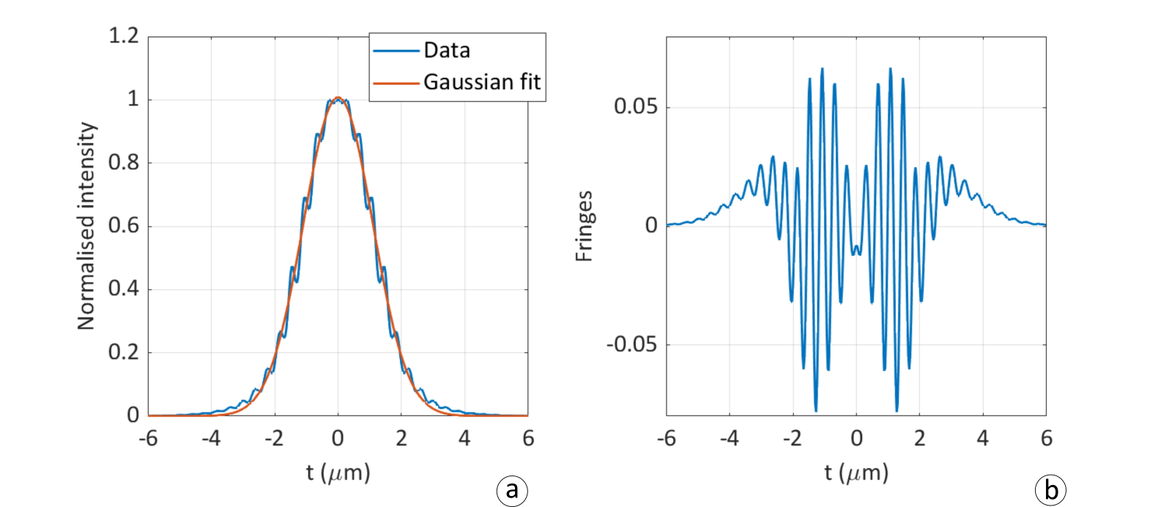}
	\label{fig4}
	\caption{a) Intensity distribution of 720 nm light in the cross region of the waveguides along transverse diagonal for two interfering modes; b) The same distribution with extracted Gaussion best fit distribution.}
\end{figure}

\begin{figure}[h]
	\centering
	\includegraphics[scale=0.6]{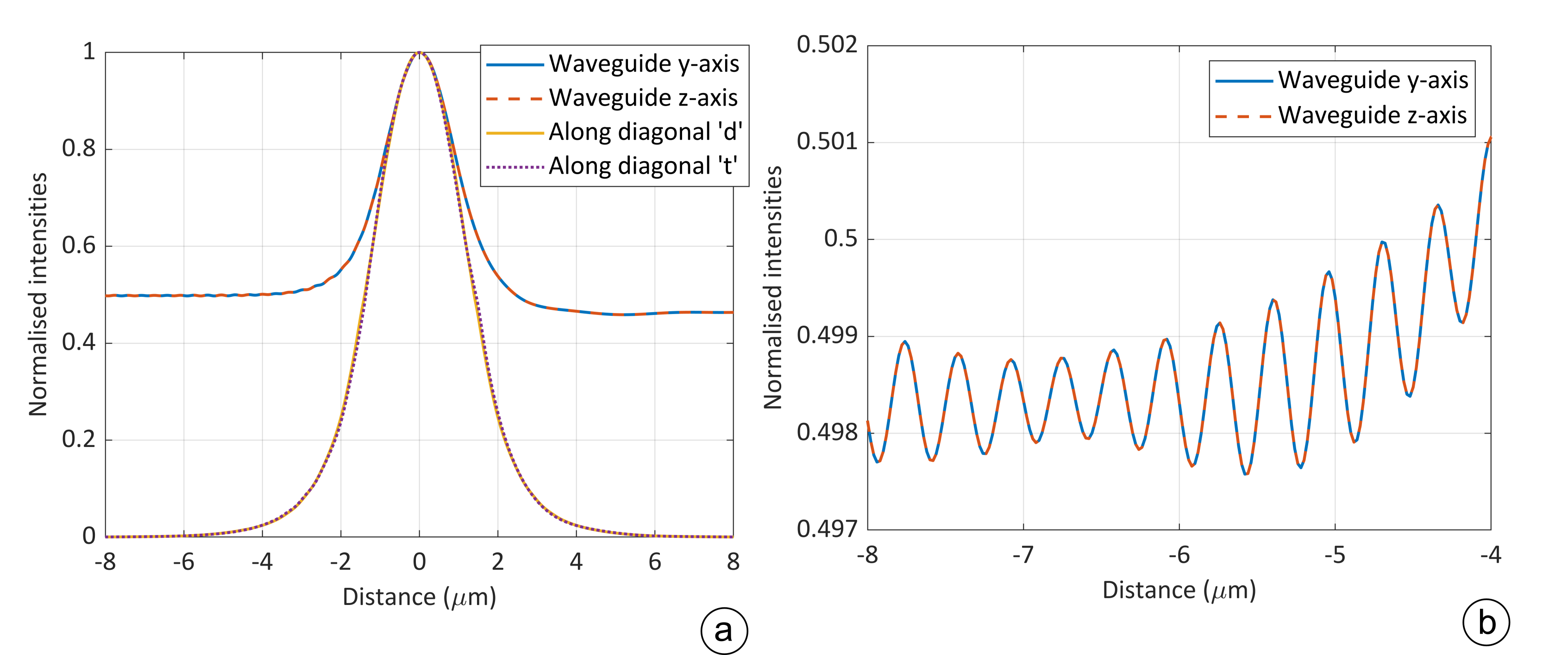}
	\label{fig5}
	\caption{a) Intensity distributions of the two crossed and non-interfering 850 nm modes of the $w_{rib}=2$ $\mu$m waveguide along different axes; b) Zoomed interference fringes along y-axis (or z-axis) caused by partial reflection of the 850 nm modes from the cross region.}
\end{figure}

The influence of the fringes on trapped atoms can be efficiently eliminated by their rapid spatial displacements, such as the time-averaged optical interference fringes being smeared out and not visible to atoms due to their inertia. There are several ways to achieve that effect. One way is to use laser light with large phase fluctuations. There are multiple cheap diode lasers, the emission of which includes multiple longitudinal modes with random fluctuations between them. Another way to achieve it is to introduce a relatively large frequency shift between the two crossed optical modes. Such a frequency shift will lead to continuous motion of the interference fringes along the t-axis and corresponding oscillating micro-motion of the trapped atoms, which is caused by the oscillating optical dipole force caused by moving interference fringes. In practice, a relative frequency shift of the order of 100 MHz, which can be produced with an acousto-optical modulator, is enough to solve the problem of interference fringes even for a Bose-Einstein condensate stored in crossed optical dipole traps.

Figure 5 shows intensity distributions of the two crossed and non-interfering 850 nm modes of the $w_{rib}=2$ $\mu$m waveguide along different axes. The top line shows the normalised intensity distributions along the y-axis and z-axis, which are identical to each other. The maximum intensity in each of the crossing optical modes is normalised to 0.5. Therefore, the maximum of two crossed modes at the centre of the cross region is about equal to 1. At the left tail of the distributions there observed periodic fringes of rather small amplitude and period of about 336 nm = $\lambda_{r1}/n_{r1}^{eff}/2$, where $\lambda_{r1}$ = 850 nm and $n_{r1}^{eff}$=1.265. These fringes can be explained by interference of the incident mode with its part, which experiences partial reflection from the cross region. This happens due to a slight difference between the effective indexes of refraction of the waveguide and its cross region with the second waveguide. Two bottom lines of the graph correspond to intensity distributions along the two diagonals of the cross region. Here is a small difference between the two curves is observed, which reflects the perturbations of modes at the cross region. Note, that even such small perturbations of the modes can cause essential perturbations of the corresponding optical potentials, which will be shown in the next section.

\section{Potentials and main parameters of crossed waveguides evanescent wave traps for rubidium atoms}

\subsection{Trapping potentials}

To calculate the three-dimensional distribution of the trapping potential near the surface of two crossed suspended optical rib waveguides we used the formula  

\begin{equation}  \label{EQ1} 
U\left(x,y,z\right) = U_{r}\left(y,z\right)\ {\mathrm{exp} \left(-{2x}/{d_r}\right)+U_{b}\left(y,z\right){{\mathrm{exp} \left(-{2x}/{d_b}\right) 
 -\frac{C_3{{\lambda }_{eff}}/{2\pi }}{x^3\left(x+{{\lambda }_{eff}}/{2\pi }\right)}}}}+mgx,
\end{equation} 

where the normalised distributions of the potential $U_r\left(y,z\right)$ and $U_b\left(y,z\right)$ at the surface of waveguides ($x$=0) were derived from numerical solutions for corresponding intensity distributions of the two fundamental modes of different colours inside the corresponding rib waveguides. Note, that each of these potentials consists of added potentials of the two crossed waveguides $I$ (directed along $y$-axis) and $II$ (directed along $z$-axis), $U_i\left(y,z\right)=U_i^I\left(y,z\right)+U_i^{II}\left(z,y\right)$, where $i=r,b$. It is assumed that there is no interference between the crossed optical modes. Local magnitudes of optical dipole potentials are proportional to corresponding local intensities of laser fields $U_i\left(y,z\right)=\alpha_i I_i\left(y,z\right)$ \cite{1}. In this paper, we considered two pairs of wavelengths of opposite frequency detunings from the main rubidium atomic transition, each of which has certain advantages and disadvantages. 
The corresponding coefficients $\alpha_i$ were calculated as in \cite{15} for the wavelength presented in Table 1. The calculation takes into account the contributions of the five closest and strongest atomic transitions.
The pair of modes with $\lambda_{b2}=640$ nm and $\lambda_{r2}=930$ nm, which are detuned by 150 nm from the main rubidium atomic transition, provide relatively small spontaneous scattering rates of photons and longest coherence times of trapped atoms.  Such large frequency detunings of the guiding optical fields also solve a problem of resonant scattering by atoms of Raman emission generated inside of the silica waveguides by the guiding light \cite{21,22,23}. On the other hand, using $\lambda_{b1}=720$ nm and $\lambda_{r1}=850$ nm modes provides higher values of optical dipole potentials and make it possible to operate CWEWTs at smaller laser powers. Although, the coherence times of the stored atoms are expected to be smaller compared to the previous case of large frequency detunings. Using $\lambda_{b2}=640$ nm and $\lambda_{r1}=850$ nm modes has its advantages, which will be discussed below. The term before the last one corresponds to attractive interaction between atoms and the dielectric surface of the substrate, which can be approximated with Casimir-Polder potential \cite{24} where $C_3$ is the Van der Waals coefficient and $\lambda_{eff}$ is the reduced atomic transition wavelength, which is providing a transition from the Van der Waals interaction at $x<<\lambda_{eff}/(2\pi)$ to the Casimir-Polder interaction at $x>>\lambda_{eff}/(2\pi)$. For rubidium atoms next a SiO$_2$ surface $C_3=5.699\times 10^{-49}$ J$\cdot$m$^3$ and $\lambda_{eff}=710$ nm \cite{24}. The last term represents a gravity force directed along the outward normal to the substrate. Although, in most cases, the gravity force did not produce much effect on the total trapping potential.

\begin{table}
	\begin{center}
		\caption{Values of $\alpha$ coefficients of optical dipole potentials for Rb$^{87}$ atoms}
		\begin{tabular}{|p{3.5cm}|p{2.4cm}|p{2.4cm}|p{2.4cm}|p{2.4cm}|}
			\hline
	wavelength (nm) & $\lambda_{b1}=720$ & $\lambda_{r1}=850$ & $\lambda_{b2}=640$ &$\lambda_{r2}=930$ \\
			\hline
 $\alpha$ ($10^{-36}$J$\cdot$m$^2$/W) & $\alpha_{b1}=4.909$ & $\alpha_{r1}=-6.616$ & $\alpha_{b2}=1.859$ & $\alpha_{r2}=-3.362$ \\
			\hline
		\end{tabular}
	\end{center}
	\label{alpha_values}
\end{table}

First a CWEWT formed by two crossed suspended optical rib waveguides with $h=300$ nm, $h_{rib}=15$ nm, $w_{rib}=2$ $\mu$m, $\lambda_{b2}=640$ nm and $\lambda_{r2}=930$ nm is calculated. Initially, we intend to get the maximum possible potential depth of such a trap for rubidium atoms by limiting the total laser power in each of the waveguides to about 30 mW. The maximum intensity of the 640 nm mode at the surface of waveguides is equal to $I_{b2}=1.65\times10^{10}$ W/m$^2$, which corresponds to the power of light in each of the two waveguides of 25.7 mW. The maximum intensity of the 930 nm mode is $I_{r2}=3.75\times10^{9}$ W/m$^2$, which corresponds to the power of light in each of the two waveguides of 6.57 mW. In Table 2 this configuration is quoted as "C1". An absolute minimum of the potential is located at $y=0$, $z=0$, $x=208$ nm and in temperature units is equal to $U_{min}^{abs}/k_B=-90.7$ $\mu$K. The absolute value of which gives the trap depth along the $x$ axis ($\Delta U_x/k_B=90.7$ $\mu$K). A vibrational frequency of atoms in this trap along the $x$-axis is equal to $\omega_x/(2\pi)=192$ kHz. The trap depth along the $y$-axis and $z$-axis is approximately twice smaller ($\Delta U_{y,z} \approx 0.5 \Delta U_x$) and approximately equal to the potential depths of each EW atomic waveguide. This is a result of adding to each other the individual potentials of the two EW atomic waveguides and taking into account small perturbations of the optical modes in the cross region. A vibrational frequency of the trap along $y$- and $z$-axis is equal to $\omega_{y,z}/(2\pi)=6.2$ kHz. An important parameter of the trap is its aspect ratio, which we define as $\gamma_{xy}=\omega_x/\omega_y=\omega_x/\omega_z$, which defines the ratio between trapping strength along the main axes of the trap. For the configuration "C1" a corresponding aspect ratio is equal to $\gamma_{xy}$(C1)$=30.8$. Figure 6a shows the contour plot of the corresponding optical dipole potential of the CWEWT in the $y0x$-plane along the centre of the cross region ($y=0$, $z=0$). Due to the axial symmetry of the CWEWT to the longitudinal diagonal $l$ (see Figure 1), the potential along $z0x$-plane is the same.  As long as this potential is a sum of the two identical guiding EW potentials of the crossed waveguides, its amplitude is twice than the corresponding potential minimums of the waveguides, which are equal to -45.25 $\mu$K. Therefore, the potential depth of the CWEWT along $y$- and $z$-axis is about -45 $\mu$K (see Figure 6b). The trap is absolutely stable along these axes since potentials in the waveguides are higher than inside the cross region, where the centre of the trap is located.  

\begin{figure}[h]
	\centering
	\includegraphics[scale=0.55]{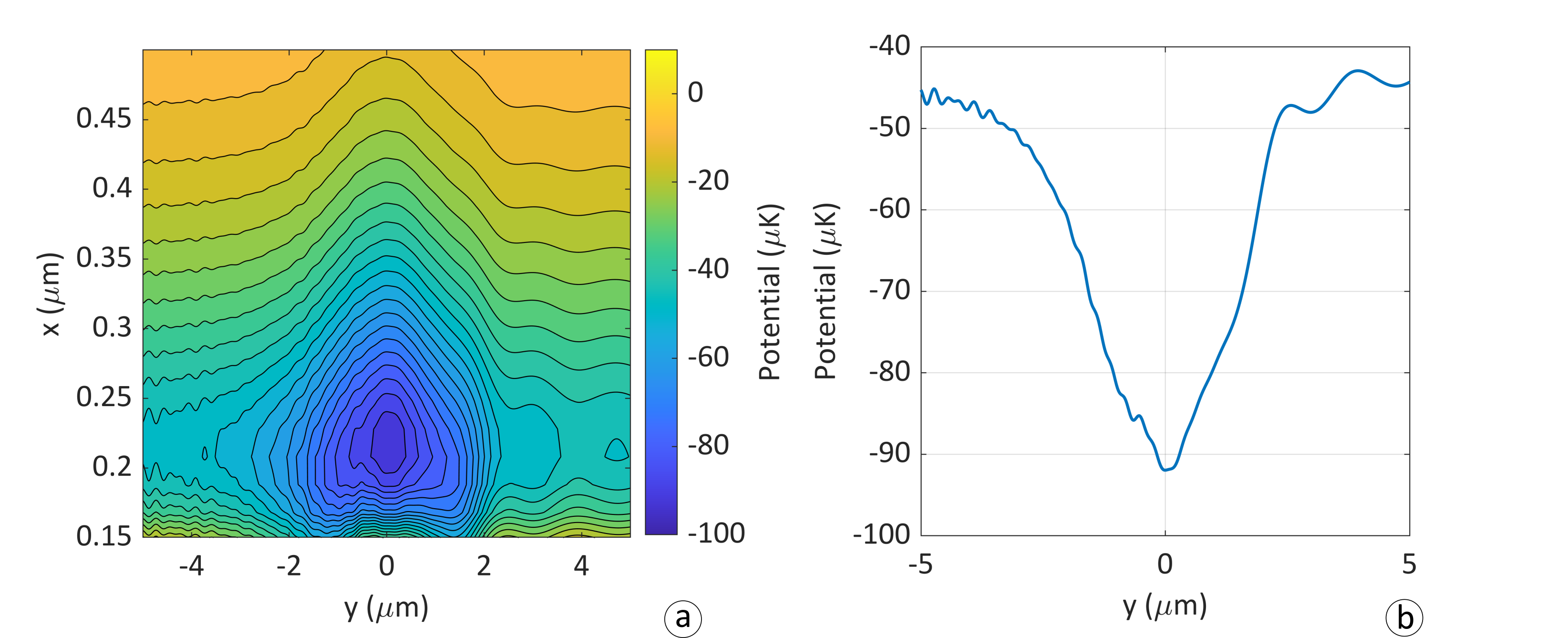}
	\label{fig6}
	\caption{a) Trapping potential of CWEWT in "C1" configuration (see Table \ref{trap_parameters}) in $y0x$-plane; b) Trapping potential along $y$-axis (or $z$-axis).}
\end{figure}

\begin{table}
	\begin{center}
		\caption{Parameters of the CWEWTs of different configurations}
		\begin{tabular}{|p{3.0cm}|p{1.7cm}|p{1.7cm}|p{1.7cm}|p{1.7cm}|}
			\hline
			Configuration & "C1" & "C2" & "C3" & "C4" \\
			\hline
			$w_{rib}$($\mu$m) & 2 & 2 & 3 & 2 \\
			\hline
			$\lambda_{b}; \lambda_{r}$(nm) & 640; 930 & 640; 930 & 640; 930 & 640; 850 \\
			\hline
			$I_{b}; I_{r}$($10^9\cdot$W/m$^2$) & 16.5; 3.75 & 6.6; 1.2 & 7; 1.2 & 4.15; 0.41 \\
			\hline
			$\Delta U_x/k_B$ ($\mu$K) & 90.7 & 21.7 & 19.5 & 6.5 \\
			\hline
			$\Delta U_{y,z}/k_B$ ($\mu$K) & 45.4 & 10.9 & 9.8 & 3.3 \\
			\hline
			$\Delta U_l/k_B$ ($\mu$K) & 12 & 2.5 & 1.06 & 2.5 \\
			\hline
			$\Gamma^{scat}_b$ (1/s) & 1.44 & 0.31 & 0.31 & 0.14 \\
			\hline
			$\Gamma^{scat}_r$ (1/s) & 0.65 & 0.67 & 0.66 & 0.5 \\
			\hline
			$\Gamma^{tun}_l$ (1/s) & 0 & <0.01 & 0 & 0 \\
			\hline
		\end{tabular}
	\end{center}
	\label{trap_parameters}
\end{table}

\begin{figure}[h]
	\centering
	\includegraphics[scale=0.7]{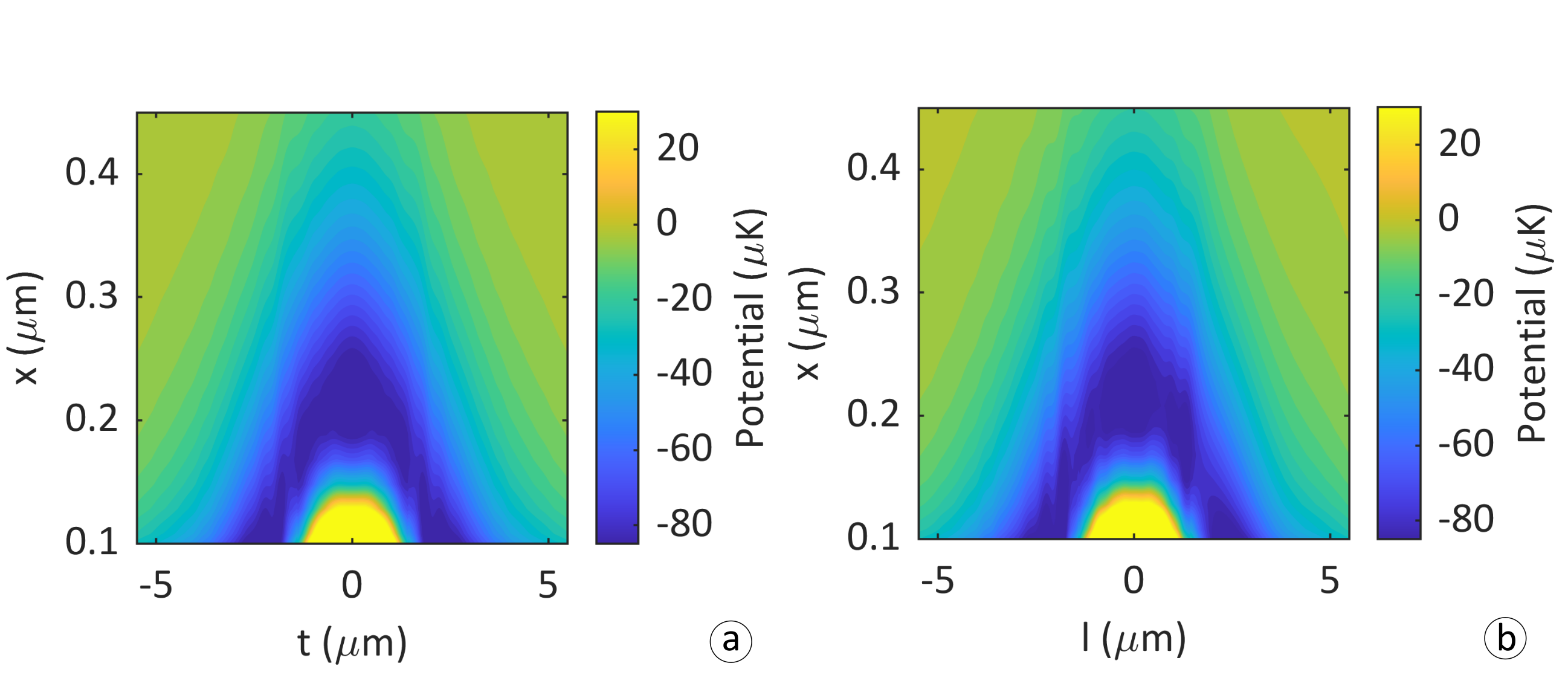}
	\label{fig7}
	\caption{Contour plot of CWEWT ("C1" configuration) trapping potential: a) in $t0x$-plane; b) in $l0x$-plane.}
\end{figure}

Figure 7 shows the cross sections of the trapping potentials in $t0x$- and $l0x$-planes, i.e. along the diagonals of the trap. These potentials look very similar to transverse potentials of the atomo-photonic waveguides \cite{15}. It can be explained by the fact that along the diagonals the intensities of the evanescent light waves are dropping to zero. At a certain critical distance from the centre of the trap, the Casimir-Polder attraction of atoms to the dielectric surface of the waveguide is getting larger than the repulsive optical dipole force of the blue-detuned evanescent light field. Therefore, these are the weakest directions of the trap, along which atoms can escape the trap by tunnelling through the optical dipole potential barrier towards the surface, as it is described in \cite{15}. Such tunnelling happens predominantly along a direction, which is characterised by the smallest potential barrier separating the trap centre from the waveguide surface. In Figure 7 this direction is characterised by a "valley" of blue colour, which is bent towards the dielectric surface.

\begin{figure}[h]
	\centering
	\includegraphics[scale=0.7]{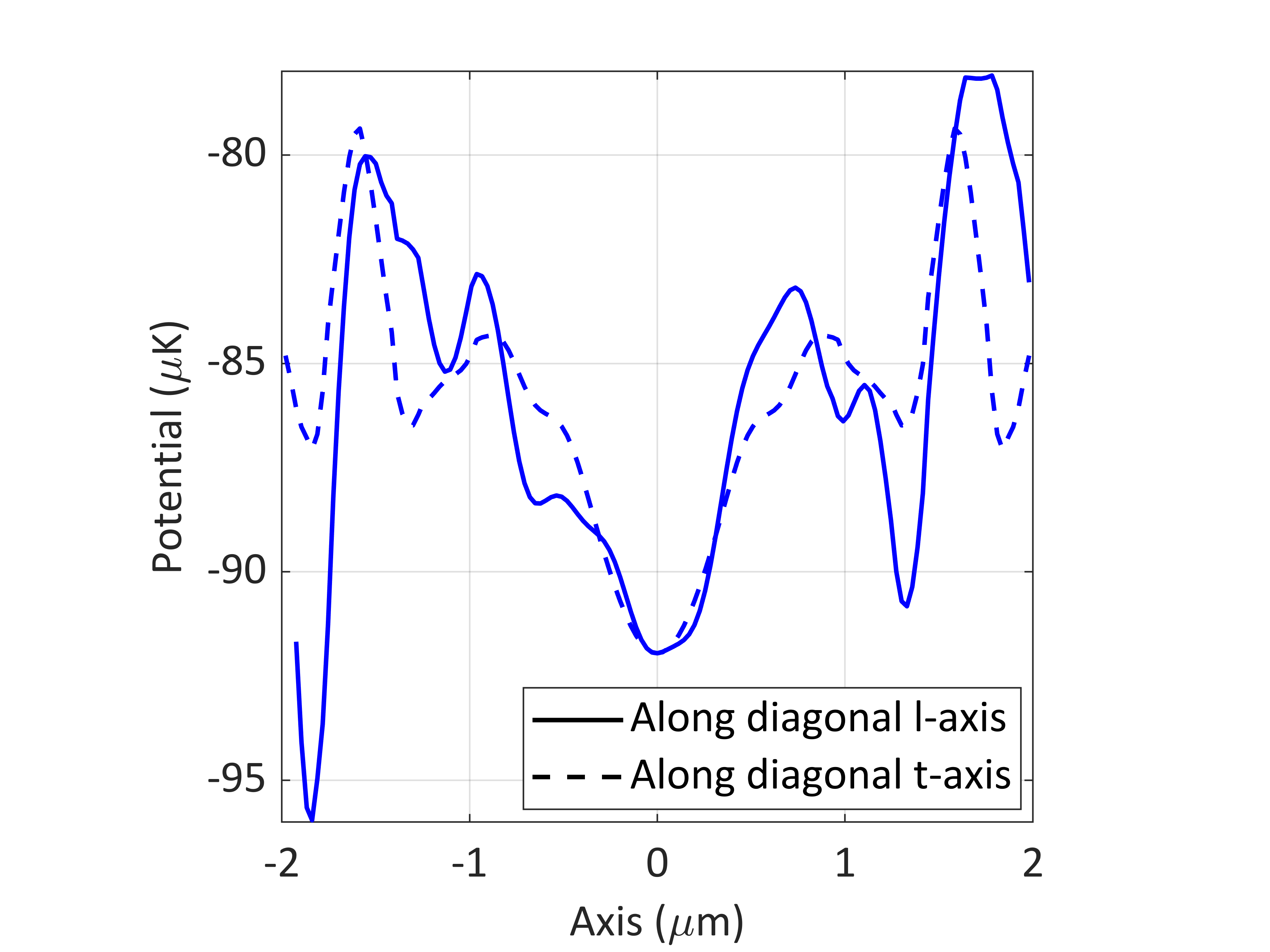}
	\label{fig8}
	\caption{Spatial distributions of CWEWT ("C1" configuration) of trapping potential local minimums along x-axis plotted along diagonal $l$-axis (solid line) and $t$-axis (dashed line).}
\end{figure}
	
Figure 8 shows corresponding minimums of the trapping potential along the $x$-axis, which are plotted as functions of $t$- and $l$-axis, i.e. transverse and longitudinal diagonals of the trap. From a symmetry of the crossed optical modes, a potential along the transverse diagonal (Figure 8a) is symmetric and along the longitudinal axis (Figure 8b) is asymmetric due to perturbations of the propagating modes in the cross region. Although the potentials along the diagonals are anharmonic, their average oscillation frequencies $\overline{\omega}_{l,t}$ are comparable to the frequencies $\omega_{y,z}$. Those average frequencies are derived from the best harmonic fits to the perturbed potentials. These plots make it possible to estimate the probability of tunnelling trapped atoms out of the trap. The probabilities of tunnelling through the potential barriers of the potential well were calculated in WKB approximation \cite{25}. The corresponding tunnelling rates of trapped atoms through the given barrier along a certain direction were estimated by multiplying the tunnelling probability by the corresponding average oscillation frequency. For a temperature of the trapped atoms of 4 $\mu$K the corresponding tunnelling rate is expected to be below $3 \times 10^{-7}$ s$^{-1}$. Note, that this is a very conservative estimate based on the one-dimensional motion of atoms along the direction of most probable escape. In Table \ref{trap_parameters} we consider all tunnelling rates, which are below $10^{-3}$ to be zero. Based on intensities of the optical fields near the minimum of the trap we estimate the spontaneous scattering rates of photons for 640 nm and 930 nm light as $\Gamma^{scat}_{b2}=1.44$ s$^{-1}$ and $\Gamma^{scat}_{r2} = 0.65$ s$^{-1}$ correspondingly. These spontaneous scattering rates of photons should provide the coherence time of the stored atoms of $\tau_{coh} = 0.48$ s.

\begin{figure}[h]
	\centering
	\includegraphics[scale=0.7]{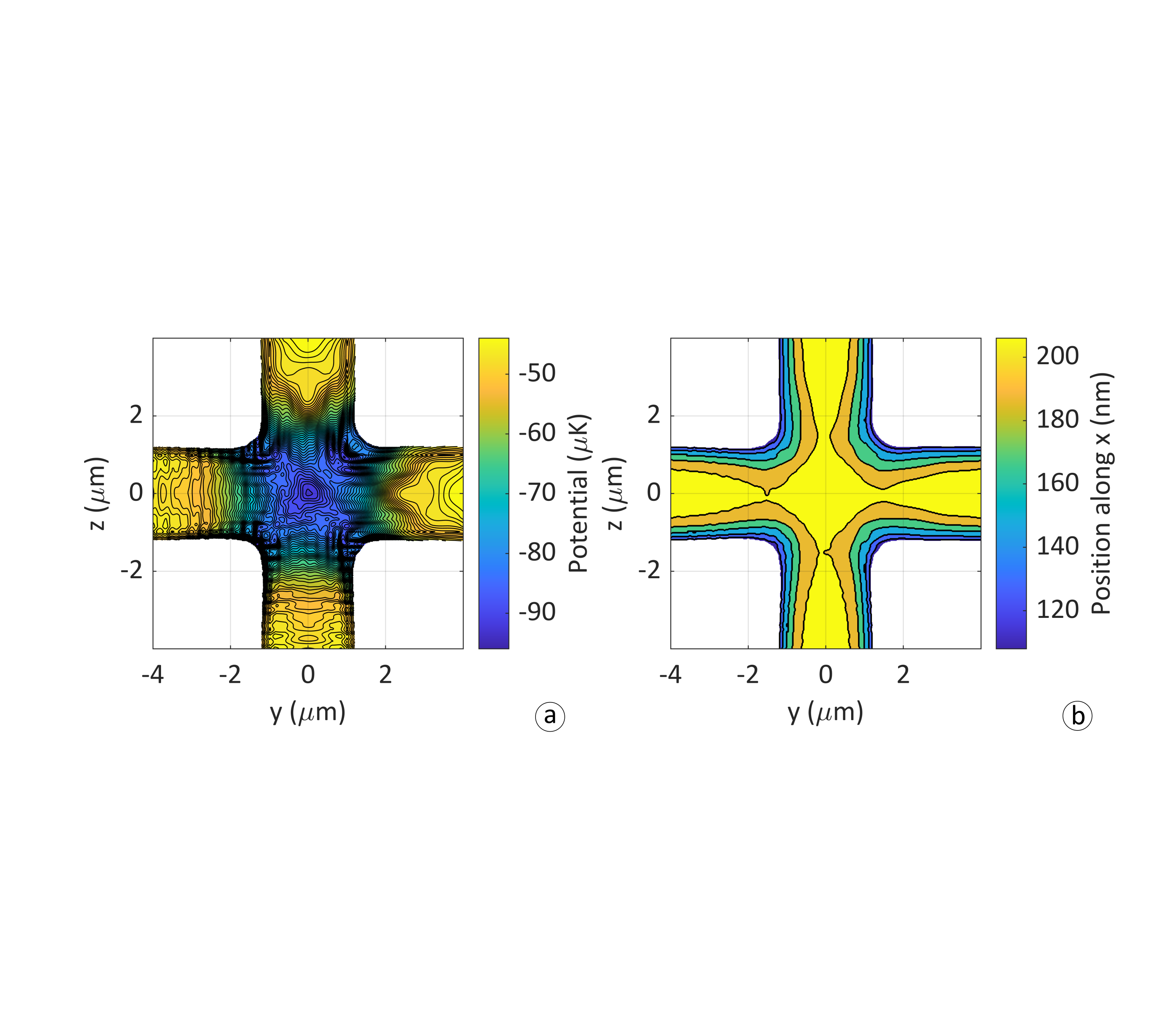}
	\label{fig9}
	\caption{a) Contour plot of the spatial distribution of CWEWT's ("C1" configuration) trapping potential minimums, calculated along the x-axis, in $y0z$-plane; b) Contour plot of distances the potential minimums from a surface of the rib in $y0z$-plane.}
\end{figure}

Figure 9a shows the 2D-spatial distribution of local trap minimums in $y0z$-plane, which were calculated along the $x$-axis. This figure illustrates that the diagonal directions of the trap ($y=z$ and $y=-z$) are the directions with the lowest potential depths. Figure 9b shows the distances of the corresponding local minimums from the surface of the rib as a function of $y$ and $z$ coordinates.

The next configuration "C2" uses the same parameters as "C1", except the intensities of 640 nm and 930 nm modes are reduced to $I_{b2}=6.6\times10^{9}$ W/m$^2$ and $I_{r2}=1.2\times10^{9}$ W/m$^2$ correspondingly. The main parameters of the trap are presented in Table 2. The minimum of the trap is located at $x_{min}=239$ nm. Its aspect ratio is $\gamma_{xy}$(C2)$=31.3$.
The smaller intensities of the evanescent light waves on the one hand lead to a smaller potential depth of the trap. On the other hand, it reduces the rates of spontaneous scattering of photons, which leads to increasing the coherence time of the trapped rubidium atoms to $\tau_{coh}=1$ s. The upper limit of the tunnelling rate of the trapped atoms with temperature $T=1.0$ $\mu$K along the longitudinal diagonal ($l$-axis) is estimated to be $\Gamma_l^{tun}<0.01$ s$^{-1}$.

\begin{figure}[h]
	\centering
	\includegraphics[scale=0.6]{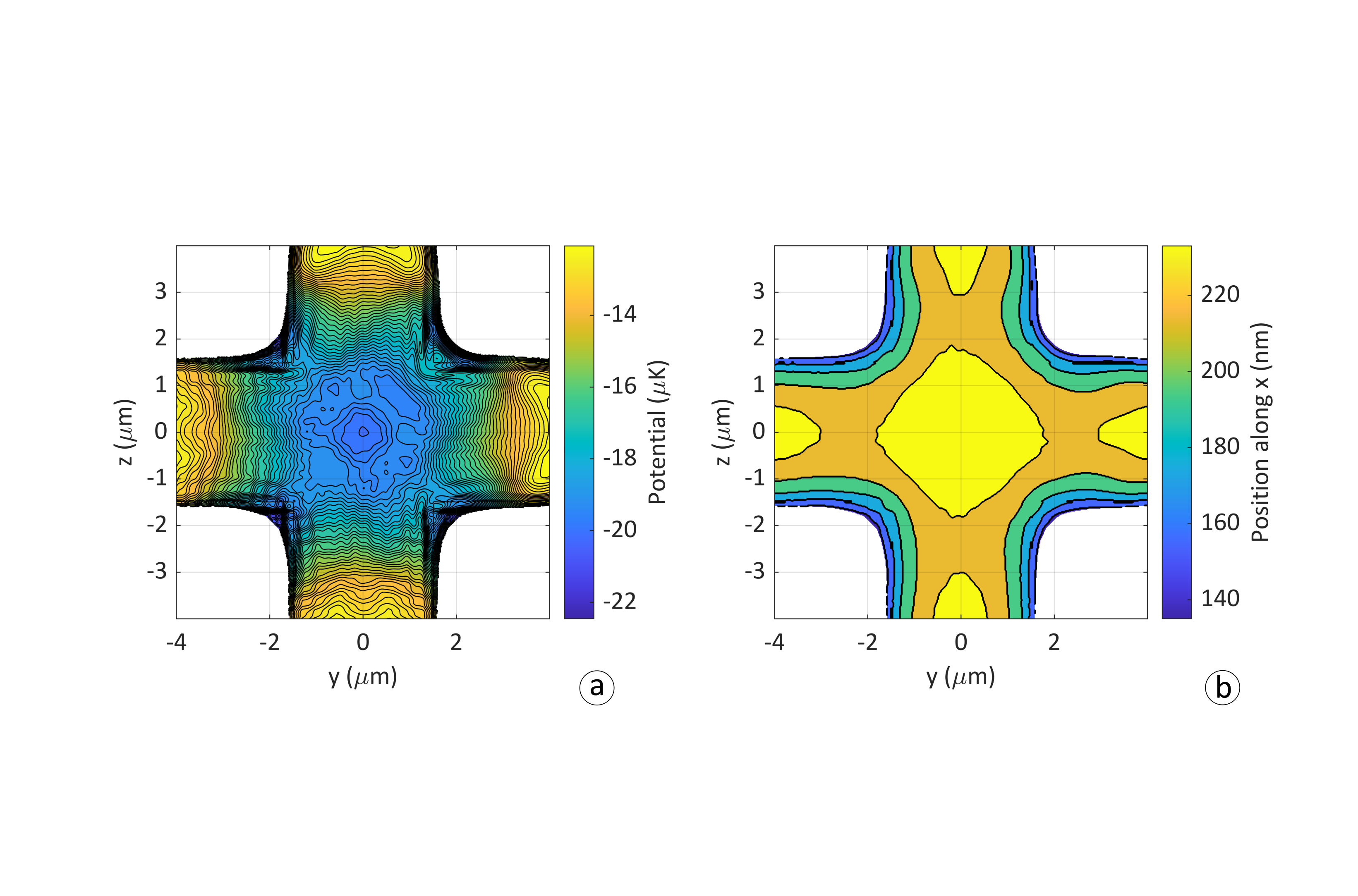}
	\label{fig10}
	\caption{a) Contour plot of CWEWT ("C3" configuration) trapping potential minimums, calculated along the x-axis, in $y0z$-plane; b) Contour plot of distances the potential minimums from a surface of the rib in $y0z$-plane.}
\end{figure}

In the configuration "C3" we used a wider suspended optical rib waveguide with rib width $w_{rib}=3$ $\mu$m, rib height $h_{rib}=5$ nm and diaphragm width $h=300$ nm. The other main parameters are shown in Table 2. One could expect that for such a wide waveguide the diffraction effects in the cross region would be smaller. Indeed the potential in Figure 10a looks more round in the centre of the cross. the vibration frequencies of the trap are $\omega_x/(2\pi)=93.9$ kHz and $\omega_{y,z}/(2\pi)=2.4$ kHz, which gives the aspect ratio of the trap $\gamma_{xy}$(C3)$=39.1$. The spatial distributions of potential minimums along the diagonal directions of the trap are sown in Figure 11. For the trapped atoms with a temperature of 0.5 $\mu$K the tunnelling rate along the longitudinal diagonal is estimated to be below $\Gamma_l^{tun}<10^{-4}$ s$^{-1}$. According to the estimated spontaneous scattering rate of photons, the coherence time of atoms in such a trap is $\tau_{coh}=1.03$ s. 

\begin{figure}[h]
	\centering
	\includegraphics[scale=0.7]{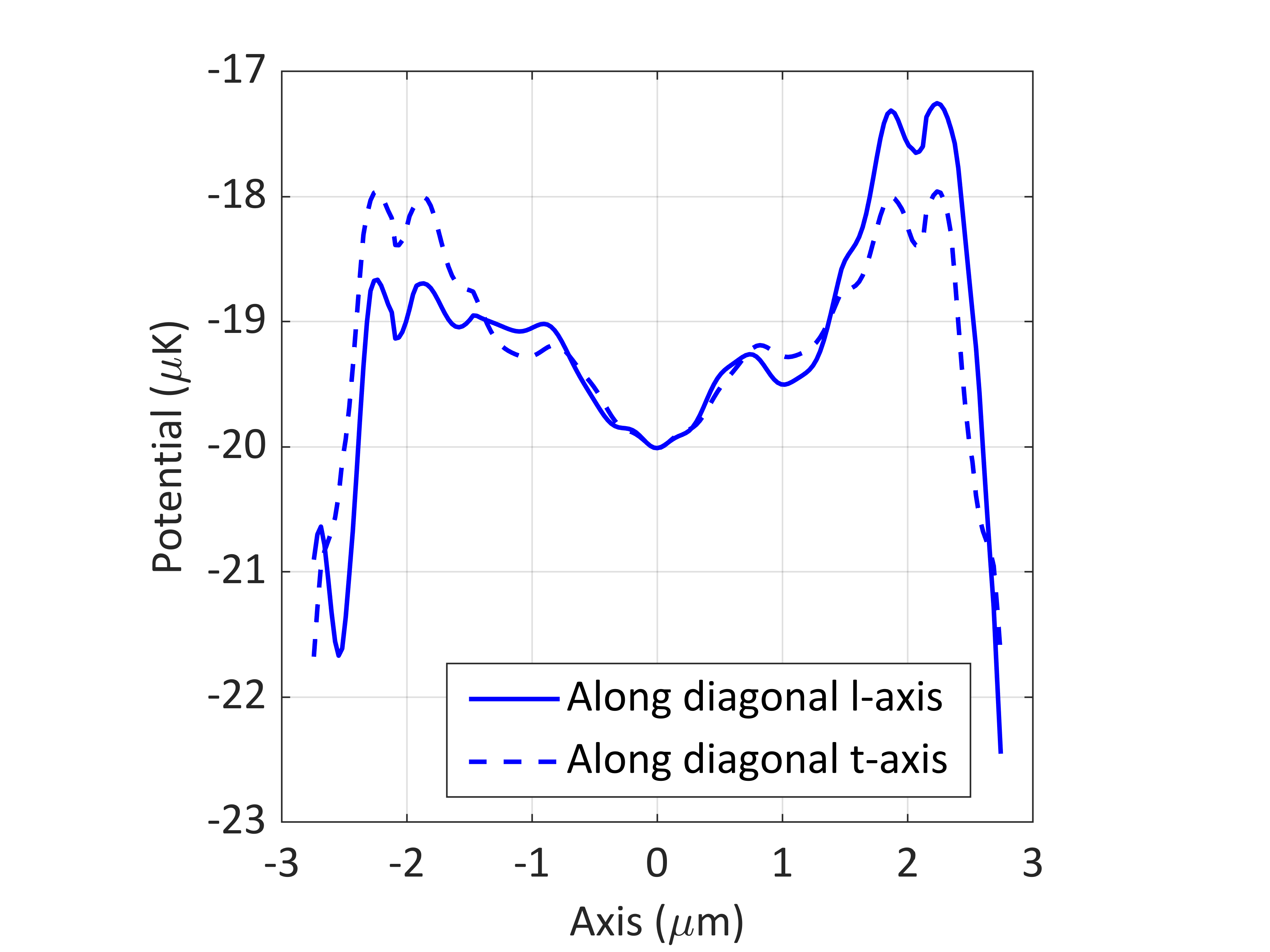}
	\label{fig11}
	\caption{Spatial distributions of CWEWT ("C3" configuration) of trapping potential local minimums along x-axis plotted along diagonal $l$-axis (solid line) and $t$-axis (dashed line).}
\end{figure}

\begin{figure}[h]
	\centering
	\includegraphics[scale=0.5]{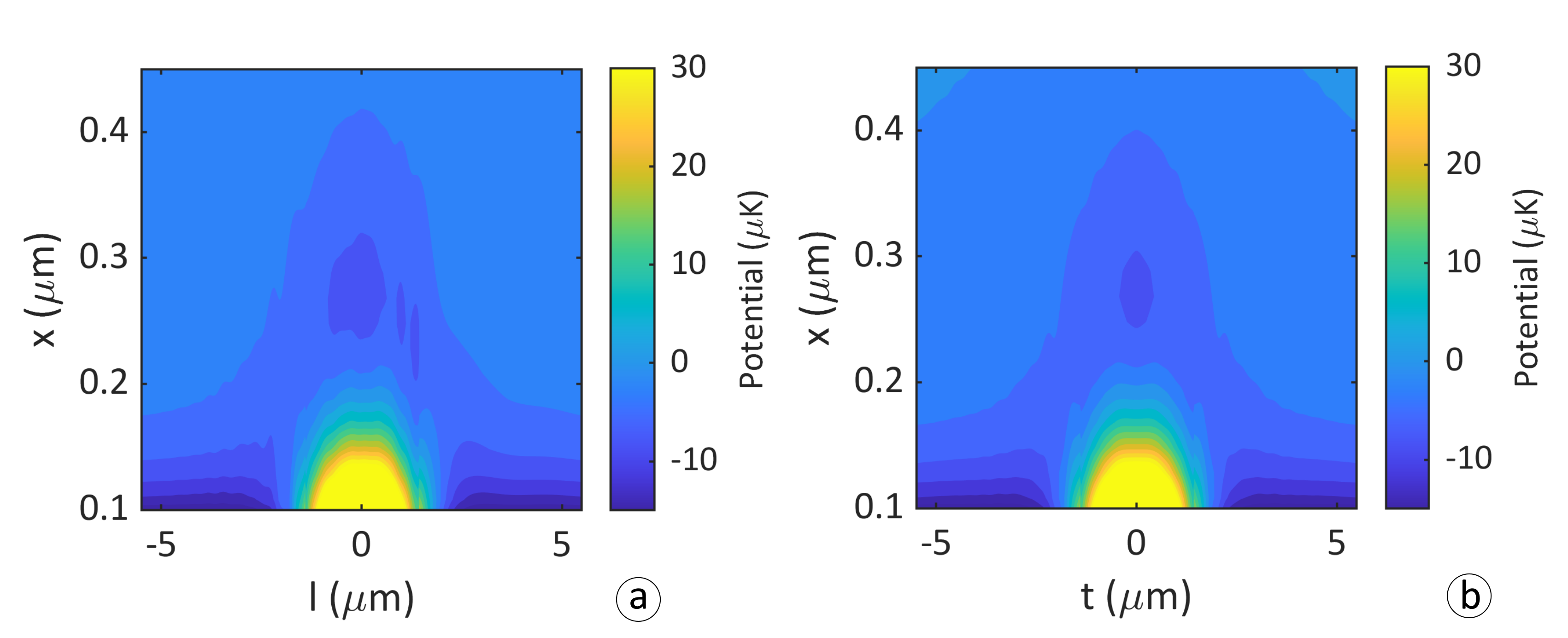}
	\label{fig12}
	\caption{Contour plot of CWEWT ("C4" configuration) trapping potential: a) in $t0x$-plane; b) in $l0x$-plane.}
\end{figure}    

\begin{figure}[h]
	\centering
	\includegraphics[scale=0.5]{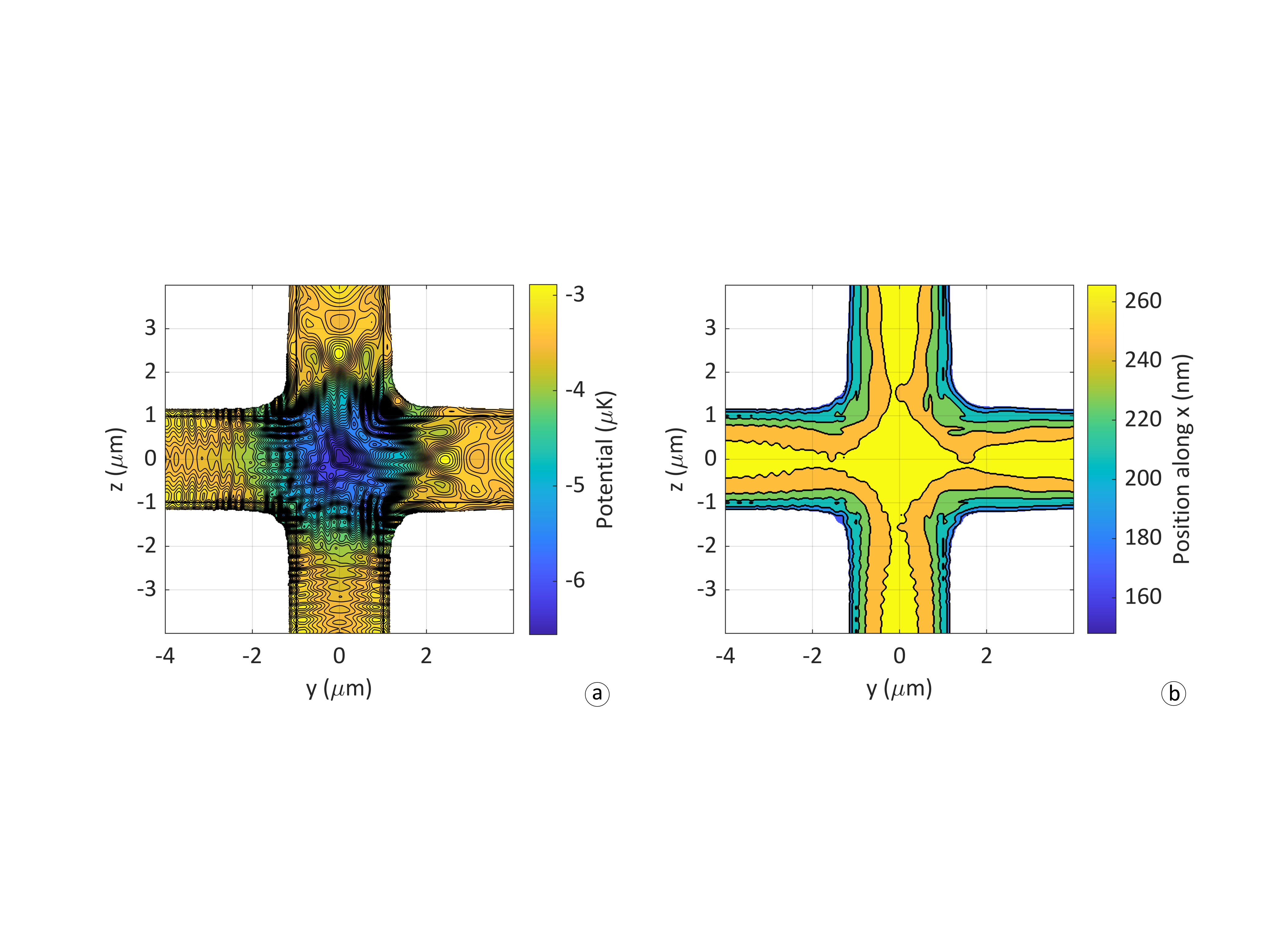}
	\label{fig13}
	\caption{a) Contour plot of CWEWT ("C4" configuration) trapping potential minimums, calculated along the x-axis, in $y0z$-plane; b) Contour plot of distances the potential minimums from a surface of the rib in $y0z$-plane.}
\end{figure}

\begin{figure}[h]
	\centering
	\includegraphics[scale=0.7]{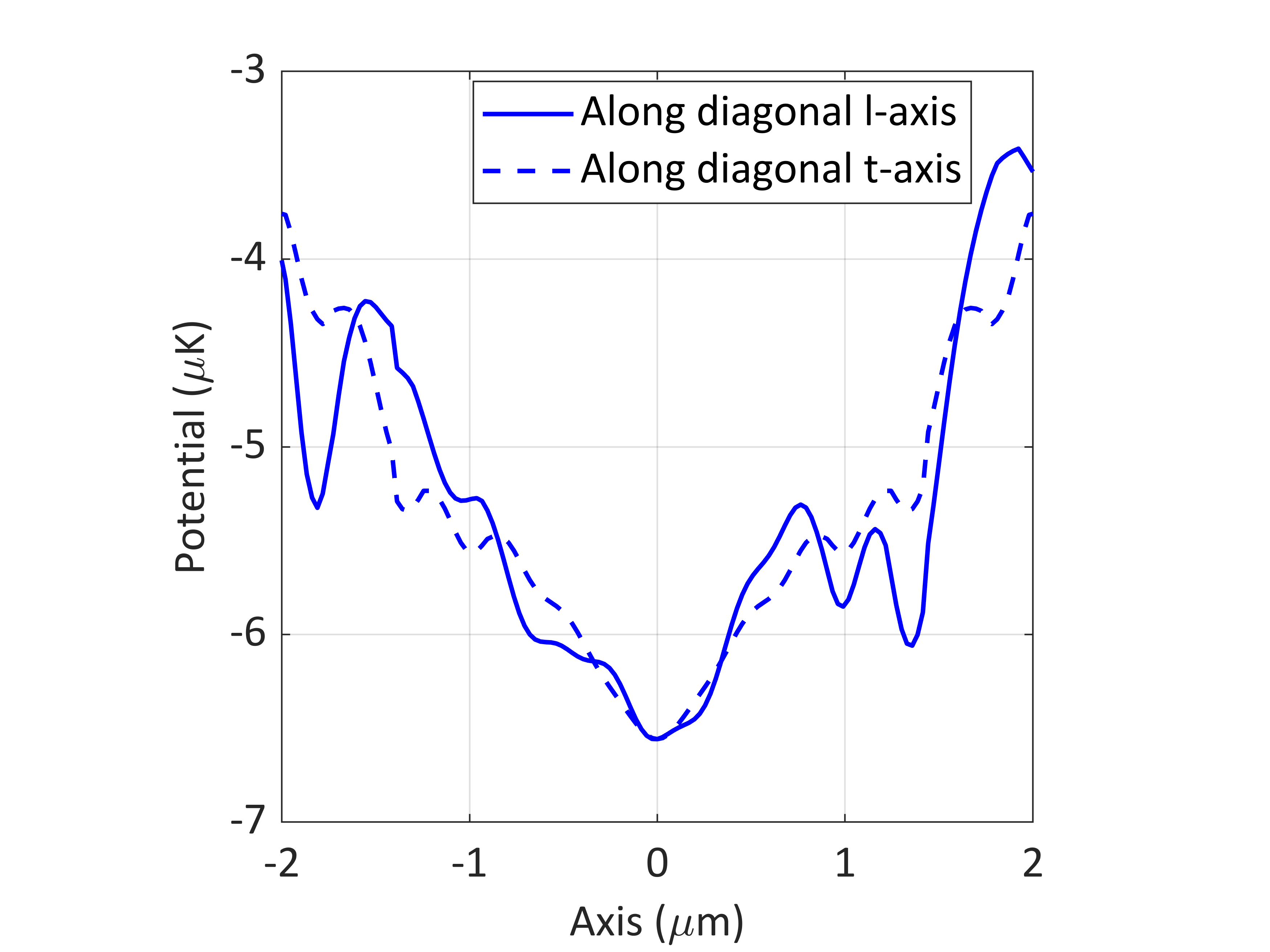}
	\label{fig14}
	\caption{Spatial distributions of CWEWT ("C4" configuration) of trapping potential local minimums along x-axis plotted along diagonal $l$-axis (solid line) and $t$-axis (dashed line).}
\end{figure}

In our final configuration "C4" we are considering the same crossed suspended optical waveguides with $w_{rib}=2$ $\mu$m, which support propagation of 640 nm and 850 nm modes. Therefore, the blue-frequency-detuned mode of 640 nm is detuned further from the rubidium atomic transition at 780 nm than the red-frequency-detuned mode of 850 nm. For the Raman scattering of these modes inside the silica waveguide, which can produce resonant light to the rubidium atomic transition, such a choice of frequencies has perfect sense as far as anti-Stokes components of such a Raman scattering are much larger than the Stokes components. In addition, the smaller frequency detuning of the 850 nm modes makes it possible to use smaller laser powers for the same potential depth of the trap. For the 640 nm modes, the intensity of the light field at the surface of the rib is $I_{b2}=4.15\cdot10^9$ W/m$^2$, which corresponds to laser power of 6.46 mW. The $I_{r1}=4.125\cdot10^8$ W/m$^2$ intensity of the 850 nm mode corresponds to a laser power of 0.2 mW. The potential of the corresponding CWEWT is shown in Figures 12, 13 and 14. It is remarkable that for the maximal depth of the trap $\Delta U_x=6.5$ $\mu$K, its minimal depth along the longitudinal diagonal is $\Delta U_l=2.5$ $\mu$K. The minimum of the trap is located at $x_{min}=262.4$ nm. Its vibration frequencies are $\omega_x/(2\pi)=54.1$ kHz and  $\omega_{y,z}/(2\pi)=2.39$ kHz. Its aspect ratio $\gamma_{xy}$(C4)$=22.7$ is smaller than in all previous configurations. The smaller intensities of the evanescent light fields of the trap lead to smaller spontaneous scattering rates and longer coherence times of stored atoms, which we estimate as $\tau_{coh}=1.56$ s. For stored atoms with temperature 1 $\mu$K, the tunnelling rate of atoms along the diagonals of the trap is estimated to be $\Gamma_l^{tun}<10^{-3}$ s$^{-1}$. Therefore, the configuration "C4" looks the best among all the four considered configurations. Its advantage can be explained mainly by large attractive optical dipole potential of the 850 nm mode, which is nearly twice stronger compared to the 930 nm mode (see Table 1). This makes it possible to operate the trap at smaller optical powers. In addition, it provides higher relative potential along the trap diagonals.

\section{Conclusion and outlook}

Integrated optical dipole traps for ultra-cold neutral atoms based on two crossed suspended optical rib waveguides are considered. It is shown that these traps are strongly anisotropic and their optical dipole potentials have two weak axes, which are directed along diagonals of the cross region. For practically achievable conditions, such traps can store ultra-cold rubidium atoms with coherence times up to 1.6 seconds. The main advantages of these optical dipole traps are their integration into a photonic chip, microscopic sizes, miniature volumes, good compression of trapped atoms in all three dimensions and ability to trap atoms in their magnetically non-sensitive internal states. These properties of the traps make them attractive for trapping single ultra-cold neutral atoms and using them as qubits. Two-dimensional arrays of such traps could be quite attractive for the realisation of integrated quantum memory devices and quantum logic elements in a similar way as it is currently done for neutral atoms trapped in arrays of optical tweezers \cite{26}. There are several potential advantages of such integrated two-dimensional arrays of qubits. First, the tight location of the trapped atoms in the CWEWTs and small distances between the adjacent qubits (about 4 $\mu$m) should provide better interaction between them, which is necessary for the generation of entangled quantum states. Second, such distances between the adjacent qubits imply individual addressability to the qubits with focussed laser beams. Finally, it might be possible to find configurations of such CWEWTs, which contain only a single vibration state, like in atomo-optical waveguides discussed in \cite{15}. Such a configuration of the CWEWTs would provide an ideal qubit, which is free of a motional decoherence of the trapped atoms.  
There are many other configurations of CWEWTs to be investigated in future.

\subsection{Acknowledgments}

We acknowledge the support of the UK government department for Science, Innovation and Technology through the UK national quantum technologies programme. Many thanks to Hugh Klein and Geoffrey Barwood for their comments and corrections to the paper.

\end{document}